\documentstyle[twoside,fleqn,espcrc2,epsf,enumerate]{article}

\newcommand{\bee}{\begin{equation}}
\newcommand{\ee}{\end{equation}}
\newcommand{\beea}{\begin{eqnarray}}
\newcommand{\eea}{\end{eqnarray}}
\newcommand{\rme}{{\rm e}}

\newcommand{\AmS}{{\protect\the\textfont2
  A\kern-.1667em\lower.5ex\hbox{M}\kern-.125emS}}

\title{Improved lattice actions}

\author{Ferenc Niedermayer\address{Institute for Theoretical Physics,
University of Bern, \\
Sidlerstrasse 5, CH-3012 Bern, Switzerland}
}
       
\begin{document}

\begin{abstract}
The main strategies to reduce lattice artifacts for spin models, 
gauge fields, free fermions and QCD are discussed.
\end{abstract}

\maketitle


\section{INTRODUCTION}

The idea to improve lattice actions \cite{Wilson2,Symanzik} 
is nearly as old as the use of lattices in regularizing 
field theories \cite{Wilson1}.
Nevertheless, until recently numerical simulations were
performed mostly with the simplest (standard) form of 
lattice actions. This attitude seems to have been altered ---
at this conference there were probably more contributions
using some kind of improved action than the standard ones.
The reason is, obviously, that the lattice artifacts resulting
from the finite lattice spacing $a$ cannot be easily reduced 
further simply by brute force, i.e. increasing the lattice size.
As explained in Ref.~\cite{Lepage} the computing cost
for full QCD grows at least as $1/a^6$. 
I would like to add to this that in some cases the situation 
is much worse, even for pure gauge theory.

\begin{itemize}

\item Thermodynamics.\\
 For measuring the basic thermodynamic 
quantities, like the free energy, a conservative estimate gives
$$
\mbox{computing cost~} \sim \frac{1}{a^{10}}.
$$ 
(This is because the free energy density, $f={\rm O}(T^4)$,
is hidden behind the ultra--violate fluctuations
of the action density,
$\langle S \rangle = {\rm O}(1/a^4)$, as will be discussed later.) 

\item Topological susceptibility.\\
Using the standard form of the action (with the geometrical
definition of the topological charge) the presence of 
dislocations --- configurations with topological charge
and too small value of the action -- spoil the result
even in the continuum limit $a\to 0$!

\end{itemize}

In this talk I will concentrate mostly on theoretical ideas
and developments, since
the numerous contributions concerning simulations with improved
actions are covered in several plenary talks at this conference.

\section{IMPROVING THE ACTIONS}

There are two main ways of improving the lattice actions, based on:
\begin{itemize}
\item
Wilson's renormalization group (RG),
\item
Symanzik's approach, inspired by perturbation theory (PT).
\end{itemize}

Wilson's approach \cite{Wilson2} is based on the fact that 
a critical system
(with a correlation length $\xi=\infty$) remains critical
after mapping the lattice to a coarser one by a block
transformation. The lattice action itself changes, in general,
under the renormalization group transformation (RGT)
and moves (except for pathological transformations as decimation)
towards a fixed point (FP) on the critical surface.
The trajectory leaving the critical surface along the relevant
(unstable) direction(s) defines the renormalized trajectory.
The system specified by an action on the renormalized trajectory
has exactly the same partition function as the original
system defined on an `infinitely fine' lattice, and describes
the same long--range physics with no lattice artifacts.
The RGT can be (to a large extent) arbitrary,
leading to different FP's and renormalized trajectories.
There is, however, an additional useful option --- one can
vary (and optimize) the RGT during the successive RG steps on
the way to smaller and smaller correlation length. 
One can call this a `running' RGT. The actions in this large set
have the property that they are {\it perfect} -- i.e. produce
no cut--off effects (for physical quantities like
masses, scattering amplitudes).

The disadvantage of this beautiful construction is that
the actions involved have complicated structure,
and it is difficult to determine and use them in numerical
simulations. Perhaps to avoid these difficulties, Symanzik
\cite{Symanzik} suggested an alternative approach.
The procedure was to add some extra terms to the naive lattice 
action and choose their coefficients appropriately to cancel 
the leading O($a^2$) lattice artifacts in the n-point 
Green--functions up to a given order in PT. 
(O($a^2$) refers to the bosonic case; for Wilson fermions 
the leading artifacts are more severe, O($a$)).
For gauge theories this requirement had to be reduced 
to `on--shell improvement', i.e. to the perturbative 
improvement of physical quantities \cite{Weisz,LW}.
The number of extra terms to be added is determined by the number
of possible ${\rm dim}=(5),6$ operators. By allowing redefinitions of
fields, however, this number can be reduced.
The number of extra operators to be added is following:

\vskip 5mm
\begin{tabular}{lccc}
              & op's    &  order    & Ref.            \\
\hline
gauge fields: & ~3-1=2  &  O($a^2$) & \cite{Weisz,LW} \\
fermions      & ~2-1=1  &  O($a$)   & \cite{SW} \\
              & 15-5=10 &  O($a^2$) & \cite{SW}
\end{tabular}
\vskip 5mm

\noindent(Some of the 10 fermionic operators contributing 
to O($a^2$) are quartic in the fermion fields.)
The statement is that to arbitrary order in $g^2$ only the
coefficients of these operators have to be determined
to cancel the corresponding cut--off dependence
of O($g^n a$) or O($g^n a^2$).
Although the Symanzik approach provides a relatively simple
modification of the action to cancel the leading artifacts,
the perturbatively determined coefficients are usually
not applicable in the case where most of the numerical
simulations are done. The other point, to which I shall return,
is that there is a large freedom of choosing the given number
of extra operators, and some additional criteria are needed
to find the choice best suited for numerical simulations.

\section{RECENT DEVELOPMENTS}

I will consider three main recent developments in the improvement
program:
\begin{itemize}
\item
non--perturbative numerical determination of O($a$) improvement
coefficients in QCD \cite{LSSW}

\item
tadpole improvement (TI) \cite{LepMac,Alford}

\item
FP actions (or classically perfect actions) 
\cite{FP2,DG_talk,AH_talk,Wiese,Burk}

\end{itemize}

The first one is a direct application of Symanzik's
improvement program for the non--perturbative situation in QCD:
instead of calculating the O($a$) cut--off effects
for some quantities in PT, the authors measure these
effects, and by tuning the coefficient $c_{SW}(g_0^2)$ of the
Sheikholeslami--Wohlert (clover) term \cite{SW}
they get rid of the leading O($a$) artifacts.
Since it is done in the framework of the full QCD,
I will discuss it at the end, and start with the simpler case
of bosonic theories.

\section{TADPOLE IMPROVEMENT}

The motivation behind this approach is the observation
that large coefficients appearing in PT come dominantly
from tadpole diagrams. Lepage and Mackenzie suggested
\cite{LepMac}
that these contributions should be removed by a mean--field
redefinition of the original fields. The corresponding
prescription for the SU(3) gauge fields is to replace
the fields in the original lattice action according to the rule
\bee
U_{\mu}(x) \to \frac{1}{u_0} U_{\mu}(x) \, ,
\label{Uredef}
\ee
where
\bee
u_0=\langle \frac{1}{3}{\rm Tr}U_{\rm plaq} \rangle^{1/4} \, .
\label{u0}
\ee
In addition, to get a better perturbative expansion, it is
suggested to use
\bee
\alpha_s = -\frac{1}{3.06839} 
\log \left( \frac{1}{3} \langle U_{\rm plaq} \rangle \right)
\ee
as an expansion parameter instead of $\alpha_{\rm latt}=g_0^2/4\pi$.

For tadpole improving (TI) a one--loop Symanzik improved (SI)
action  (where the coefficients are $g$ dependent) the prescription 
is to recalculate the improvement coefficients for the 
modified action obtained by (\ref{Uredef}) using ordinary PT, 
substitute $\alpha_{\rm latt} \to \alpha_{\rm s}$, and use 
this action in the numerical situation. 
The procedure involves a self--consistent determination of the
coefficients in the action: the coefficients to be used in the
MC simulation have to be determined from MC simulations.
Technically, this is not a serious problem, since it is very
easy to measure $\langle U_{\rm plaq} \rangle$.
(Principally, however, the mean field definition can cause
some problems. For example, $\langle U_{\rm plaq} \rangle$
depends not only on $\beta$ but also on the size of the system.
Should one change $u_0$ with the temperature at a given $\beta$
according to eq.~(\ref{u0}) or should it be determined
from the zero temperature, infinite volume case? 
One can argue both ways.)

I cite here two examples of tadpole improvement.
For the O($a$) one--loop Symanzik--improved Wilson fermions
the coefficient \cite{Wohlert,Naik,LW2}
\bee
c_{SW} = 1 + 0.2659 g_0^2 + \ldots
\label{cSW}
\ee
should be compared with the TI
tree--level Symanzik--improved action which gives
$$
c_{SW}^{(0)TI} = 
\left( \frac{1}{3} {\rm Re Tr} \langle U_{\rm plaq} 
\rangle \right)^{-3/4} = 1 + 0.25 g_0^2 + \ldots
$$
In other words, 94\% of the original one--loop correction is removed
by making the TI in the tree--level formula.
Accordingly, tadpole improving the 1-loop formula amounts to
using $c_{SW}^{TI}= 1 + 4\pi\cdot 0.0159 \alpha_s$.
For the pure gauge field TI for the 1-loop SI action
of L\"uscher and Weisz \cite{LW} (who have used the 
plaquette (pl) and the rectangle (rt) ) means
$$
\frac{\beta_{\rm rt}}{\beta_{\rm pl}}=
-\frac{1}{20}(1+2.0146\alpha_s) \to 
-\frac{1}{20u_0^2}(1+0.4805\alpha_s)
$$
i.e. 76\% of the coefficient has been removed
by applying (\ref{Uredef}) to the tree--level result.

The TI action for gauge fields has been demonstrated to
work well even for lattices as coarse as $a\sim 0.4{\rm fm}$
\cite{Lepage}. This success and the simplicity of the prescription
resulted in a large number of contributions to this conference 
which apply TI actions in numerical simulations.
However, as we shall discuss later, TI does not work everywhere,
e.g. for full QCD it does not give properly the value of 
$c_{SW}$ at the $\beta$ values of interest \cite{LSSW}.

After discussing the FP actions, I will try to suggest
how to understand the tadpole improvement from the corresponding 
point of view.

\section{FIXED POINT ACTIONS}

To be specific, I will consider here the pure gauge theory.
A RGT is defined by specifying the kernel of the block transformation,
$T(V,U)$. Here $V$ is a gauge configuration on the coarse
lattice while $U$ denotes a configuration on the fine lattice.
Given the action $S(U)$ one defines $S'(V)$ by
\bee
\rme^{-\beta'S'(V)}=\int dU \rme^{-\beta [S(U)+T(V,U)]} \, .
\label{rgt}
\ee
($S'(V)$, as well as $S(U)$, is normalized to the continuum
expression for very smooth fields.)
$T(V,U)$ should satisfy a normalization condition to leave
the partition function unchanged.
Since the theory is asymptotically free, for $\beta\to\infty$
one has $\beta' = \beta+{\rm O}(1)$, and in this limit
eq.~(\ref{rgt}) reduces to a saddle point equation:
\bee
S'(V)=\min_U \left\{ S(U) + T(V,U) \right\} \, .
\ee
The fixed point of this transformation is given by
\bee
S^{\rm FP}(V)=\min_U \left\{ S^{\rm FP}(U) + T(V,U) \right\} \, .
\label{fpeq}
\ee
Note, that although the $\beta\to\infty$ limit has been taken,
the fields $V$ and $U$ are not assumed to be necessarily smooth
 --- they could be arbitrarily rough as well in this definition.
Eq.~(\ref{fpeq}) can be solved analytically for smooth
configurations and numerically for arbitrary $V$ by iteration.
The iterative procedure converges very rapidly since the typical
action density on the fine lattice is $\sim 30$ times smaller than
that on the coarse lattice (a factor 16 coming from the volume,
and typically half of the value of the r.h.s. is due to $T(V,U)$.)
Consequently, in most cases one can use on the r.h.s. just the
quadratic approximation to $S^{\rm FP}(U)$ or even the standard
action.

The FP action has remarkable properties --- it defines
a {\it classically perfect} lattice action \cite{FP}, 
in the following sense. 
The solutions to the lattice equations of motion are related to 
their continuum counterparts, with the same value of the action:
\bee
S_{\rm cont}(U_{\rm cont}) = S_{\rm latt}^{\rm FP}(U_{\rm latt}) \, .
\ee
In particular, for continuum theories possessing instantons,
the FP action has scale invariant instanton solutions as well.
One can say that the FP action is `tree--level Symanzik--improved 
to all orders in $a$'. 
By proper choice of $T(V,U)$ (which is achieved by tuning
some free parameters in it) $S^{\rm FP}$ can be made quite
short ranged --- the couplings outside the unit hypercube
can be suppressed (relative to the plaquette) by a factor $\sim 100$
\cite{FP2}. 
The FP action is not expected to be a perfect action 
--- using it in numerical simulations with finite $\beta$ 
(at finite correlation length) some cut--off dependence
remains. Numerical simulations for $2d$
spin models \cite{FP,BBHN,Burk} and $4d$ SU(3)
gauge theory \cite{FP2} show, however, that the lattice 
artifacts using FP actions are negligible for these theories 
even on extremely coarse lattices.   

\begin{figure}[htb]
\vspace{5pt}
\begin{center}
\leavevmode
\epsfxsize=60mm
\hspace{20pt}
\epsfbox{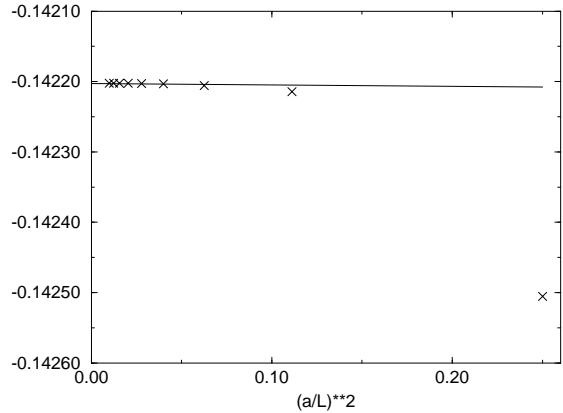}
\end{center}
\vspace{-10pt}
\caption[]{The value of $R$
vs. $(a/L)^2$ for the FP action for $L/a = 2, 3, \ldots, 10$. 
The fit shown is $ -0.1422022 - 0.0000176\, (a/L)^2$. 
Note that the exact limiting value is $-0.1422024$.}
\label{fig:1lperf}
\end{figure}

Actions which are (quantum) perfect are obtained by 
following the renormalized trajectory (also allowing
to vary $T(V,U)$ with $\beta$).
Strong arguments suggest that the FP action is also
1--loop perfect, i.e.
\bee
\beta S^{\rm perf}(U;\beta) =
\beta S^{\rm FP}(U) \left[ 1 + {\rm O}\left( \beta^{-2} \right) \right]
 \, .
\ee
This conjecture is supported by formal RG arguments
\cite{Wilson2,FP2} and by numerical evidence for the O(3)
non--linear sigma--model \cite{FHNP}. 
In  \cite{FHNP} it was found that the mass gap in a small box
calculated to 1--loop in PT with the FP action has no O($g^4a^n$)
error. 
A coefficient $R$ appearing in the 1--loop expansion of the 
mass gap is plotted vs. $(a/L)^2$ in Fig.~\ref{fig:1lperf}.
As seen, the correction is vanishing as $\exp(-\gamma L/a)$.
This reflects the requirement that the size of the box
should be larger than the interaction range of the FP action
(otherwise the infinite--volume action should be modified).
The tiny power--like artifact seen in the fit is caused by
restricting the interaction distances considered 
in the perturbative calculation.

Although the FP action is classically and presumably 1--loop perfect,
the approximation actually used in MC simulations
contains some {\it imperfect} steps:
\begin{itemize}
\item
truncation in the interaction distance,
\item
parametrization (choice of the set of operators to fit the numerically
known values of the FP action)
\item 
the FP action is not on the renormalized trajectory (it is not 
`quantum perfect').
\end{itemize}
One has to control these points to get a reliable FP action.
It produces a misleading result, for example, to restrict the RGT
to a small subspace of operators beforehand.
One must show first that the interaction range is not larger
than the truncation radius for the interaction terms
and truncate the set of operators only afterwards.

\subsection{Natural choice of improvement terms}
In Symanzik's approach one has a large freedom to choose
the improvement terms. For the tree--level improvement to 
${\rm O}(a^2)$ one can choose the following loops, for example, 
\begin{enumerate}[A.)]
\item
the plaquette and the rectangle ($2\times 1$ loop),
\item
the plaquette and the $2 \times 2$ loop.
\end{enumerate}
Which one should be preferred?
Let me illustrate the point on the $2d$ $\sigma$--model.
Denote by $\rho_{xy}$ the coupling between two spins
at the relative distance $(x,y)$. (The standard action
is given by the nearest--neighbour  term $\rho_{10}=1$.)
Take two choices of operators in the Symanzik approach:
\begin{enumerate}[a.)]
\item
$\rho_{10}=4/3$, $\rho_{20}=-1/12$, 
\item
$\rho_{10}=2/3$, $\rho_{11}=1/6$, 
\end{enumerate}
The FP values for these coefficients (for an `optimal' RGT
\cite{FP}):
$$
\rho_{10}=0.618,~ \rho_{11}=0.190 ,~ \rho_{20}=0.002
$$
The choice a.) has an antiferromagnetic coupling and 
a complex dispersion relation in the quadratic approximation. 
Clearly, our RGT prefers the choice b.) in the Symanzik program,
which is free from these diseases.

Similarly, for gauge theory the FP action in quadratic approximation
gives a much smaller coefficient for the rectangle (rt) than
for the `bent rectangle' (brt) or the parallelogram (pg) \cite{FP2}. 
(Note that the brt and pg are equivalent in the quadratic
approximation.) In the standard tree--level SI form, however,
the plaquette (pl) and the rt are the two operators chosen 
to have non--zero coefficients. This choice does not correspond
to the FP action. In quadratic approximation it produces again
a complex spectrum above some momenta.
In fact, L\"uscher and Weisz \cite{LW} proposed a one--parameter
family of SI actions with a free coefficient $x$ for the brt,
suggesting that $x\ne 0$ could be useful in MC simulations.
However, as far as I know, only the $x=0$ choice has been exploited.

I think, for the Symanzik program a natural choice of the improvement 
operators could be the one inspired by the RG approach. 
One can also turn the argument around. 
As mentioned above, the FP action is automatically Symanzik--improved,
to all orders in $a$.
A necessary truncation to a reasonable number of operators
does not keep, of course, this property. Nevertheless, for a 
FP action with a short interaction range the violations of the lowest 
order (in $a$) Symanzik conditions will be small. 
Since the truncation is to some extent an arbitrary procedure, 
one might also keep these conditions satisfied. 
I note here, that we were interested mainly in a good parametrization
of the FP action at relatively rough configurations (typical at short
correlation length) hence we did not always impose the above conditions.

\subsection{Parametrization of the FP action}

In \cite{FP2} the FP action has been parametrized as
\bee
S^{\rm FP}(U)=\sum_{\cal C} \sum_m c_m({\cal C}) X_{\cal C}^m \,,
\label{Sparam}
\ee
with
\bee
X_{\cal C}={\rm ReTr}\left( 1-U_{\cal C} \right) \,,
\ee
and ${\cal C}$ denotes a closed loop. A few simple loops were
taken with their powers $m=1..4$. The higher powers are essential
in this parametrization, and they do not cause much overhead
in numerical simulations.
With a Swendsen-type blocking in ref.~\cite{FP2} a good scaling
was observed for $a\le 0.4{\rm fm}$. In addition, a good
asymptotic scaling for $T_c/\Lambda$ has been found, which could
indicate that the bare PT for the FP action works better.

On the renormalized trajectory the coefficients in eq.~(\ref{Sparam})
will be $g$ dependent. The 1--loop perfectness, however, predicts
that this dependence is given by
\bee
c_m({\cal C},g)=c_m({\cal C}) + {\rm O}(g^4)\,,
\ee
i.e. no O($g^2$) corrections are expected. 
One concludes that no large tadpole contributions could
be present for the FP action, and it would be {\it incorrect}
to tadpole improve it by
\bee
S^{\rm FP}(U) \to S^{\rm FP}(U/u_0)\,.
\ee
On the other hand, a mean field approximation to the FP action
(\ref{Sparam}) yields an action of the TI type:
\bee
S^{\rm FP} \approx \sum_{\cal C} 
c_{\rm eff}({\cal C},\langle X_{\cal C}\rangle ) X_{\cal C}\,,
\ee
where
$
c_{\rm eff}({\cal C},\langle X_{\cal C}\rangle )=
c_1({\cal C})+ 2 c_2({\cal C}) \langle X_{\cal C}\rangle +\ldots
$
Indeed, the relative weight of the parallelogram (pg)
to the plaquette, $c_{\rm eff}({\rm pg})/c_{\rm eff}({\rm pl})$
grows with $g^2$ as suggested by the tadpole improvement
(where it is $\propto 1/u_0^2$).

\subsection{Improving correlators of local operators; 
the $q\bar{q}$ potential}

\begin{figure}[htb]
\vspace{10pt}
\begin{center}
\leavevmode
\epsfxsize=60mm
\hspace{-20pt}
\epsfbox[100 80 450 400]{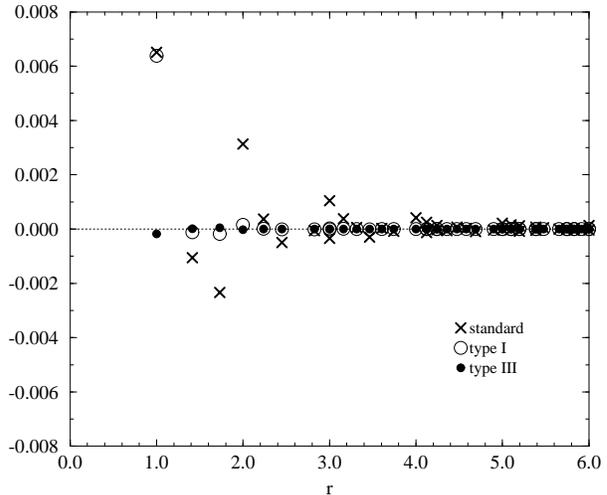}
\vspace{0pt}
\end{center}
\vspace{-10pt}
\caption[]{The difference $V(r)-V_{\rm cont}(r)$ vs. $r$ for the
Wilson action, type--1 and type--3 FP actions.}
\label{fig:pertpot}
\end{figure}

\begin{figure}[htb]
\vspace{-5pt}
\begin{center}
\leavevmode
\epsfxsize=60mm
\hspace{-30pt}
\epsfbox[100 80 450 430]{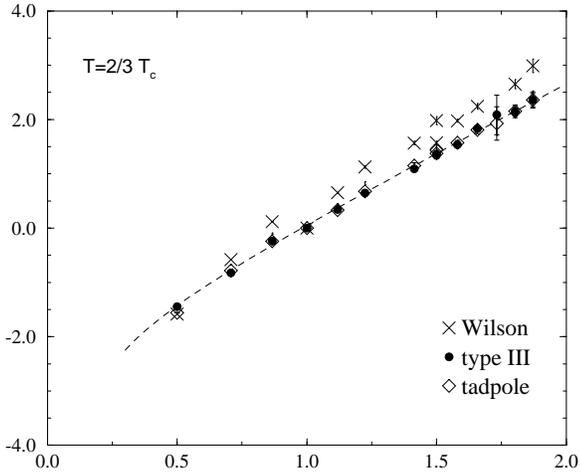}
\vspace{0pt}
\end{center}
\vspace{-30pt}
\caption[]{$V(r;T)/T_c$ vs. $r T_c$ for the
Wilson, type--3 FP and 1--loop TI actions, at finite temperature,
$T=2/3 \cdot T_c$. The dashed line is the universal curve.}
\label{fig:pot_3tw}
\end{figure}

In the Symanzik approach one improves the action to cancel cut--off
effects in spectral quantities. To achieve this for correlation 
functions one also has to improve the operators. The situation
is the same in the RG approach. For a square--shaped
blocking, although the spectrum is exact, the potential
shows violations of rotational symmetry. (A close analogy:
in the continuum electrodynamics the interaction between 
square--shaped charges.)
The perturbative $q\bar{q}$ lattice potential has, in general,
terms of the following type:
$$
V(r)\sim \frac{1}{r} +\ldots \frac{a^2}{r^3}
+\ldots\frac{a^4}{r^5}{\rm P}_4(\cos\theta) 
+\ldots\rme^{-\gamma r/a} \,.
$$
The second term (related to the distorted gluon spectrum) is removed 
by improving the action, the third one (an octupole term) by improving
the operator. In the RG approach, the last term comes from the 
interaction of the quantum fluctuations on the fine lattice for a
given coarse configuration (and is similar in nature to the
exponentially vanishing term in Fig.~\ref{fig:1lperf}).
In Ref.~\cite{FP2} using a Swendsen--type RGT (our type--1 RGT)
we have got a $q\bar{q}$ potential which gave the proper string
tension at $a=0.4{\rm fm}$ but had considerable violation of
rotational symmetry at small distances. This has been associated 
in ref.~\cite{FP2} with the strongly square--shaped averaging of 
the RGT applied. To illustrate this point, and also to obtain 
an even better FP action, in Ref.~\cite{BN}
a more general RGT (type--3) has been considered, which included 
generalized staples along the $2d$ and $3d$ diagonals. 
As expected, the rotational invariance at small distances has been 
improved both for the perturbative and the non--perturbative potential.
In Fig.~\ref{fig:pertpot} the deviation of the perturbative potential
from the continuum result is shown. In Fig.~\ref{fig:pot_3tw}
the $q\bar{q}$ potential measured at finite temperature 
$T=2/3\cdot T_c$ and $a\approx 0.4{\rm fm}$ is compared for
the Wilson action and type--3 FP action \cite{BN}. For comparison, I
measured also the 1--loop tadpole--improved potential for this
temperature, also shown in  Fig.~\ref{fig:pot_3tw}.
Surprizingly, the tadpole and the type--3 FP data almost coincide.

Although it is not related to the FP actions, I end this section 
with a reference to \cite{Garcia}
where it has been suggested to test improvement schemes by using analytic
and numerical methods in a small physical volume.
The authors proposed a combination of the 
plaquette, the $2\times 1$  and the $2\times 2$ loops,
with coefficients $16/9$, $-1/9$ and $1/144$ which also satisfy the
tree--level Symanzik conditions.
The improvement for the small volume quantities is about the same as
for the L\"uscher--Weisz choice ($5/3$, $-1/12$ and $0$). 
The choice of \cite{Garcia} is motivated by the observation that their
analytic calculations could be extended to investigate the effects
of the TI in this case.

Let me also mention a contribution \cite{Morn} where the 
authors used the 1--loop SI + TI gauge action on {\it anisotropic}
lattices to calculate the glueball mass. The use of an anisotropic
grid which is finer in the time direction ($a_t<a_s$) is a useful tool
to study heavy objects. The data for the $0^{++}$ and $2^{++}$
glueballs scale well up to $a_s=0.4{\rm fm}$, while the standard 
action shows large artifacts on coarse lattices. 

\section{THERMODYNAMICS}

Thermodynamic quantities measured at relatively large temperatures 
are sensitive not only to the low energy part of the spectrum
hence the lattice artifacts in this case are usually much larger.
For example, in PT for $T\to\infty$ at temporal lattice size $N_t=4$
the cut--off effect for $p/T^4$ using Wilson's gauge action is 
$\approx 50\%$ \cite{Beinlich}.

The pressure is given by 
\bee
\left. \frac{p}{T^4} \right|_{\beta_0}^{\beta}
= N_t^4 \int_{\beta_0}^{\beta} d\beta'
\left( \langle S\rangle_0 - \langle S\rangle_T \right) \,.
\ee
Here the action densities on the r.h.s. are measured
at a given $\beta'$ for two different temporal geometries,
at $N_t=\infty$ (or rather for a 4--cube $N_t=N_s$) and $N_t$
corresponding to the given temperature $T$.
(Actually, this is the formula which shows that the corresponding
computer cost grows like $N_t^{10}$, as mentioned in the
introduction.)
In Ref.~\cite{Beinlich} the authors measured $p/T^4$ with the 
standard action for $N_t=4,6,8$ and a few improved actions, 
including the $2\times 1$ 1-loop TI action for $N_t=4$. 
(For the results see the figure in Ref.~\cite{Karsch}.) 
Papa ~\cite{Papa} has measured the same quantity for the FP 
actions with RGT of type--1 and type--3, for $N_t=2,3$ and $T=2T_c$.
The relative deviations from the continuum result (obtained by extrapolating
the standard action results) for  $T=2T_c$ are given below.

\vskip 3mm
\begin{tabular}{ccccc}
 $N_t$ & Wilson  & type--1 & type--3 & tadpole \\
\hline
   2   & 100\%   &  25\%   &   10\%  &         \\
   3   &         &  10\%   &  "0\%"  &         \\
   4   &  20\%   &         &         &  "0\%"  \\
\hline
\end{tabular}
\vskip 3mm
\noindent{}"0\%" means that no deviation is seen within the error.
Note that the $N_t=2$ case is special because of possibly large
$\exp(-\gamma N_t)$ type corrections.

One can conclude that the cut--off effects are drastically reduced
using these improved actions.

\section{TOPOLOGICAL SUSCEPTIBILITY}

The FP action is especially suited for measuring topological effects:
\begin{itemize}
\item
it has scale invariant instanton solutions (for sizes $\rho \ge 0.7a$)
\item
a good definition of topological charge is possible (classically
perfect $Q$)
\end{itemize}

\noindent{\bf Definition:} Measure the topological charge on the 
{\it minimized} fine configuration $U=\overline{U}(V)$ of
eq.~(\ref{fpeq}),
\bee
\overline{Q}(V) = Q(\overline{U}(V))
\label{fpQ}
\ee
(Here I mean a minimization iterated a few times if necessary, on even
finer grids. Since on the fine grids the fields are rather smooth, 
after the second iteration the value of $Q$ usually does not change.)
The charge $\overline{Q}(V)$ avoids the problem of dislocations:
any configuration $V$ with $\overline{Q}(V)=1$ has a value of the
FP action not smaller than the continuum result,
$S^{\rm FP}(V) \ge S_{\rm cont}^{\rm inst}$.

By measuring the topological charge this way, one also avoids
problems associated with the cooling procedure (e.g. the disappearance
of the instantons) --- the minimization
procedure provides a prescription to separate the fluctuations from
a smooth background solution.
The configuration $\overline{U}(V)$ is smoother than $V$, it is
closer to a solution --- part of the energy of the fluctuations
in $V$ is taken over by $T(V,U)$ in eq.~(\ref{fpeq}). 
If $V$ is a solution then $T(V,\overline{U}(V))=0$.
Consequently, the minimization $V\to \overline{U}(V)$ provides
an ideal cooling procedure.
A serious drawback is that the minimization procedure is very slow,
it is hard to use it in a MC measurement. A solution is to find
an approximate analytical form for the function $\overline{U}(V)$
as has been done for the spin models \cite{BBHN,Burk}.

This strategy has been applied to the 2d non--linear sigma--model
\cite{BBHN}, the 2d ${\rm CP}^3$ model\cite{Burk}, 
and to the 4d SU(2) gauge model \cite{DHZ}. 
For the O(3) model $\chi_t \xi^2$  does not scale
(not unexpectedly) while the ${\rm CP}^3$ model shows a beautiful
scaling, for $\xi\ge 7$. In addition, asymptotic scaling is found
in $m/\Lambda_{\rm 2-loop}$ for $\xi \ge 4$, suggesting again that
the bare PT should work better for the FP action.
Of course, in the $4d$ SU(2) gauge theory the technical difficulties 
are more serious, and the investigations are not completed yet.
Nevertheless, the results for the scaling of the topological
susceptibility \cite{DHZ} are encouraging.
For the corresponding plots see Ref.'s~\cite{Burk,DHZ}.

Let me discuss a few points in some detail.
Blocking down a smooth instanton solution to a coarser lattice 
remains an instanton solution until $\rho \sim 0.7a$ where
$\rho$ is the size of the instanton. Instantons of smaller size
`fall through the lattice' --- they cannot be distinguished from
local excitations. This is the consequence of the classical 
(saddle point) approximation. 
By fixed coarse configuration $V$ at finite $\beta$ in eq.~(\ref{rgt}) 
the typical fine configuration $U$ fluctuate around
the minimizing configuration $\overline{U}(V)$.
These can have different charges, especially if the action density
of $V$ is localized to a small region.
In contrast, $\overline{U}(V)$ assigns a unique fine configuration
to $V$, and for instantons which `fell through the lattice'
this gives zero topological charge.
Hence the contribution of these small instantons is lost 
by using the FP action and FP topological charge on the coarse lattice. 
Nevertheless, the definition (\ref{fpQ}) of the FP--charge associates
$\overline{Q}=0$ for these small size excitations so they don't spoil
the topological susceptibility. One expects to observe scaling of
$\chi_t \xi^d$ once the role of instantons of size
$\rho\le 0.7a$ becomes negligible. By using the standard action
and even the (tree-level or 1--loop) SI action, 
there is a wide range of instanton sizes
where the instanton action is smaller than the continuum value
while the charge associated is 1 in the geometrical definition
--- see the corresponding figure in Ref.\cite{Burk}.
As a consequence, one can grossly overestimate $\chi_t$ and observe
scaling only at much finer lattices. I also have the feeling that
using TI will not help much in this case --- the mean field
approximation represents the action at some average `roughness'
of the fields, while instantons of different sizes are perhaps
important at a given value of $\beta$.
In Ref.~\cite{Wiese} the authors discuss the notion of the
`quantum perfect topological charge' on the example
of the exactly soluble 1d XY--model.

Note that the use a parametrized form for the minimized fine field
$\overline{U}(V)$ technically might be similar to applying 
appropriately smeared fields to measure the topological 
charge \cite{DiGiacomo}.

\section{NON--PERTURBATIVE DETER\-MI\-NA\-TION OF IMPROVEMENT 
COEFFICIENTS}

In QCD with Wilson fermions the Wilson term (which has been introduced
to avoid the doubling problem) explicitly breaks chiral invariance 
and induces ${\rm O}(a)$ cut--off effects in chiral relations.
To remove these artifacts it has been proposed to modify the
fermion action by a Pauli term 
$a c_{\rm SW}\bar{\psi}\sigma_{\mu\nu}\psi F_{\mu\nu}$.
The lattice representation of this operator is chosen to be the
`clover' term \cite{SW}.
The coefficient $c_{\rm SW}(g_0)$ has been determined in the SI 
approach to 1--loop order and given by eq.~(\ref{cSW}) 
\cite{Wohlert,LW}.
In Ref.~\cite{LSSW} the dependence of this coefficient on the
coupling $g_0$ has been determined by non--perturbative methods.
The framework used was the Schr\"odinger functional method, where
instead of periodic boundary conditions in the time
direction Dirichlet b.c.'s are used for the gauge fields and fermions.
This has the advantage that a background field can be introduced in
the system and the response to this field can be investigated.
By using a small physical box and Dirichlet b.c. the zero modes in the
fermion matrix are avoided, hence this framework is especially suited
to study the chiral limit $m\to 0$.

\begin{figure}[htb]
\vspace{-15pt}
\begin{center}
\hspace{-30pt}
\leavevmode
\epsfxsize=80mm
\epsfbox{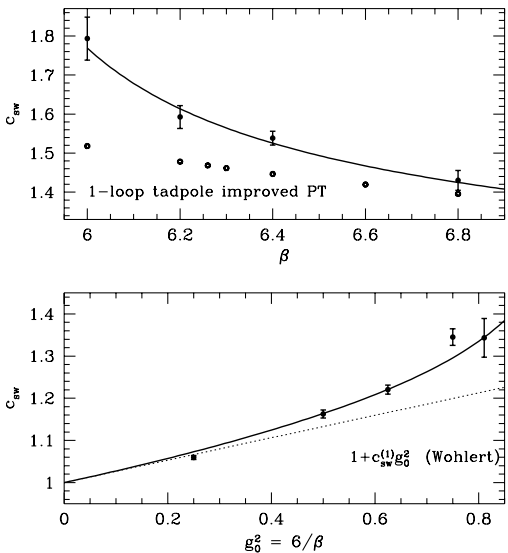}
\end{center}
\vspace{-50pt}
\caption[]{Dependence of $c_{\rm SW}$ on the coupling,}
\label{fig:cSW}
\end{figure}

From the PCAC relation $\partial_{\mu} A_{\mu}^a(x)=2mP^a(x)$
for the axial current and pseudoscalar density, valid in 
the continuum limit, the authors define
\bee
m=\frac{ \langle \partial_{\mu} A_{\mu}^a(x) {\cal O}^a \rangle }
{ \langle P^a(x) {\cal O}^a \rangle } \,,
\label{mdef}
\ee
where ${\cal O}^a$ is a fermionic operator on the boundary.
For Wilson fermions this quantity depends very strongly on the
position $x$ and the background field, even at $a=0.05{\rm fm}$!
Introducing the SW term into the action, and an extra term in 
the current via
\bee
{\cal A}_{\mu}^a(x) = A_{\mu}^a(x) + c_A a \partial_{\mu} P^a(x) \,,
\ee
by properly tuning the coefficients $c_{\rm SW}$ and $c_A$
one can make the quantity (\ref{mdef})
independent of $x$ and the background field. Fig.~\ref{fig:cSW}
shows the dependence of $c_{\rm SW}$ on the coupling \cite{LSSW}.

We can conclude that the SW choice for the
improvement term works well, but the coefficient is strongly
non--perturbative in the interesting region of couplings.
Perhaps one can try other schemes where the higher order 
(or non--perturbative) corrections are smaller.
Note also that the tadpole improvement of the 1--loop SI result
at $\beta=6$ fails by about 15\% and the discrepancy seems to
grow rapidly with $g_0$.

In a work presented at this conference \cite{Goeckeler} 
G\"ockeler et al. used the value $c_{\rm SW}$ predicted 
by the ALPHA collaboration \cite{LSSW}. Their results at $\beta=6$
shown on an APE plot seem to extrapolate well to the physical
${\rm N}/\rho$ ratio.
R. Gupta noted that minimizing the source--dependence in his
measurements prefers a value of $c_{\rm SW}(g_0)$ consistent 
with the ALPHA result \cite{Gupta}.

\section{FERMIONS}

For fermions, compared to pure gauge fields, the cut--off effects
are larger and the computational cost grows faster with 
the lattice size, hence it is especially important to optimize 
the choice of the lattice action used.

\subsection{D234$\left( \frac{2}{3}\right)$ action}

In their contribution to this conference, Alford, Klassen and Lepage
\cite{Alford} suggested a modification of the `D234' fermionic action
\cite{D234} which they call `D234$\left( \frac{2}{3}\right)$'. 
It is constructed using higher order lattice derivatives 
along the axis.
In addition the authors also introduce anisotropic lattices, 
by taking different lattice spacings along the spatial and 
time directions, $a_s/a_t \sim 2~-~3$. 
With this additional trick the dispersion
relation for free fermions becomes real and reasonably good.
(For the isotropic lattice the dispersion relation is complex
above $a p \approx 1$.) Note that the anisotropic choice,
besides curing the spectrum has the advantage that one can measure
time correlations on a finer scale which can be important for heavy
objects.
The authors have plotted the `velocity of light' $c({\bf p})$, 
defined via $E({\bf p})^2 = E(0) + c({\bf p})^2 {\bf p}^2$, 
measured for different actions in the quenched approximation 
for $\pi$ and $\rho$. 
While the TI of 1--loop SI gluon+fermion action gives a large
deviation from 1, the  `D234$\left( \frac{2}{3}\right)$'
action on an anisotropic lattice produces good agreement
with $c({\bf p})=1$. 

I would like to point out that choosing an anisotropic lattice
can be a useful tool in itself for dealing with heavy objects, 
but it is not needed to improve the spectrum of free fermions.
As investigations of the FP fermionic actions show this goal can be
achieved by using appropriate couplings on the unit hypercube.

\subsection{FP actions for fermions}

The RG approach to the fermion problem is presented at this conference
 by two groups, from MIT \cite{Wiese} and 
from Bern/Boulder \cite{DG_talk}.
The approaches of the two groups are very similar,
the main difference is in the type of the blocking transformation
used.  This difference is of technical nature, the results
obtained and the expectations are very similar.
 
The FP action for the massless free fermions has the form
\bee
S_{\rm F}^{\rm FP}=
\sum_{n,r} \bar{\psi}(n) \left[ \gamma_{\mu} \rho_{\mu}(r) +
 \lambda(r) \right] \psi(n+r) \,.
\label{fermfp}
\ee
The features of this FP actions are:
\begin{itemize}
\item
The spectrum is exact, there are no ${\rm O}(a^n)$ cut--off 
effects (`classically perfect')
\item
There are no doublers, but chiral symmetry is violated. This
violation is, however, very special: it comes entirely from 
the blocking transformation which is not affecting the physics.
\item
By optimizing the corresponding RGT the interaction range could be
made quite small --- couplings decrease exponentially fast and
outside the unit hypercube could perhaps be neglected.
\end{itemize}

The $m\ne 0$ case is very similar, only in this case one does not
have a FP, of coarse, since the mass is running. It is possible to
start with an arbitrarily small mass and make an appropriate number
of blocking transformations, doubling the mass in each step,
until the given value of $ma$ is reached. An important point here
is the following: the RGT which was optimal for the $m=0$ case
will not be optimal for the massive case --- by applying a fixed RGT
the interaction range would grow. 
To stay with a short ranged action one has to modify
the parameters of the RGT with the growing mass. This can be done
for all RGT's considered, and one can keep the action short ranged
--- at least up to $ma \approx 2$ but perhaps even higher.

In numerical simulation one needs to truncate the interaction distance
$r$ in eq.~(\ref{fermfp}). This introduces some imperfectness:
the spectrum is distorted, `ghosts' (higher lying and sometimes
complex branches in the spectrum) appear,
Symanzik's improvement conditions for the spectrum will be
(slightly) violated. Nevertheless, since by the truncation one neglects
only small coefficients, these effects are not expected to be large.
Table~\ref{tab:fp} lists some of the coefficients of a FP action
(corresponding to one of the Bern/Boulder RGT's) for free massless
fermions. The coefficients $\lambda(r)$ and $\rho_1(r)$ decrease
rapidly with the distance $r$ hence truncation to the unit hypercube
seems to be justified.

\begin{table}
\begin{center}
\begin{tabular}{|c|c|c|}
\hline
$r$  & $\lambda(r)$ & $\rho_1(r)$ \\
\hline
0000 & $~2.312461$  &             \\ 
1000 & $-0.112098$  &  $0.140881$ \\
1100 & $-0.034557$  &  $0.034424$ \\
1110 & $-0.014912$  &  $0.010742$ \\
1111 & $-0.007373$  &  $0.003714$ \\
2000 & $~0.001493$  &  $0.001380$ \\
3000 & $-0.000088$  &  $0.000243$ \\
4000 & $~0.000000$  &  $0.000010$ \\
\hline
\end{tabular}
\end{center}
\caption{Some coefficients of a fermionic FP action.}
\label{tab:fp}
\end{table}

\begin{figure}[htb]
\vspace{9pt}
\epsfxsize=62mm
\hspace{6mm}
\epsfbox{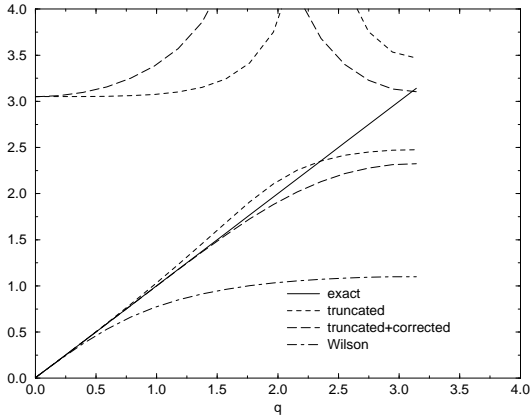}
\vspace{-10pt}
\caption{Fermion spectrum at $ma=0.005$.}
\label{fig:sp0005}
\end{figure}

\begin{figure}[htb]
\vspace{9pt}
\epsfxsize=62mm
\hspace{6mm}
\epsfbox{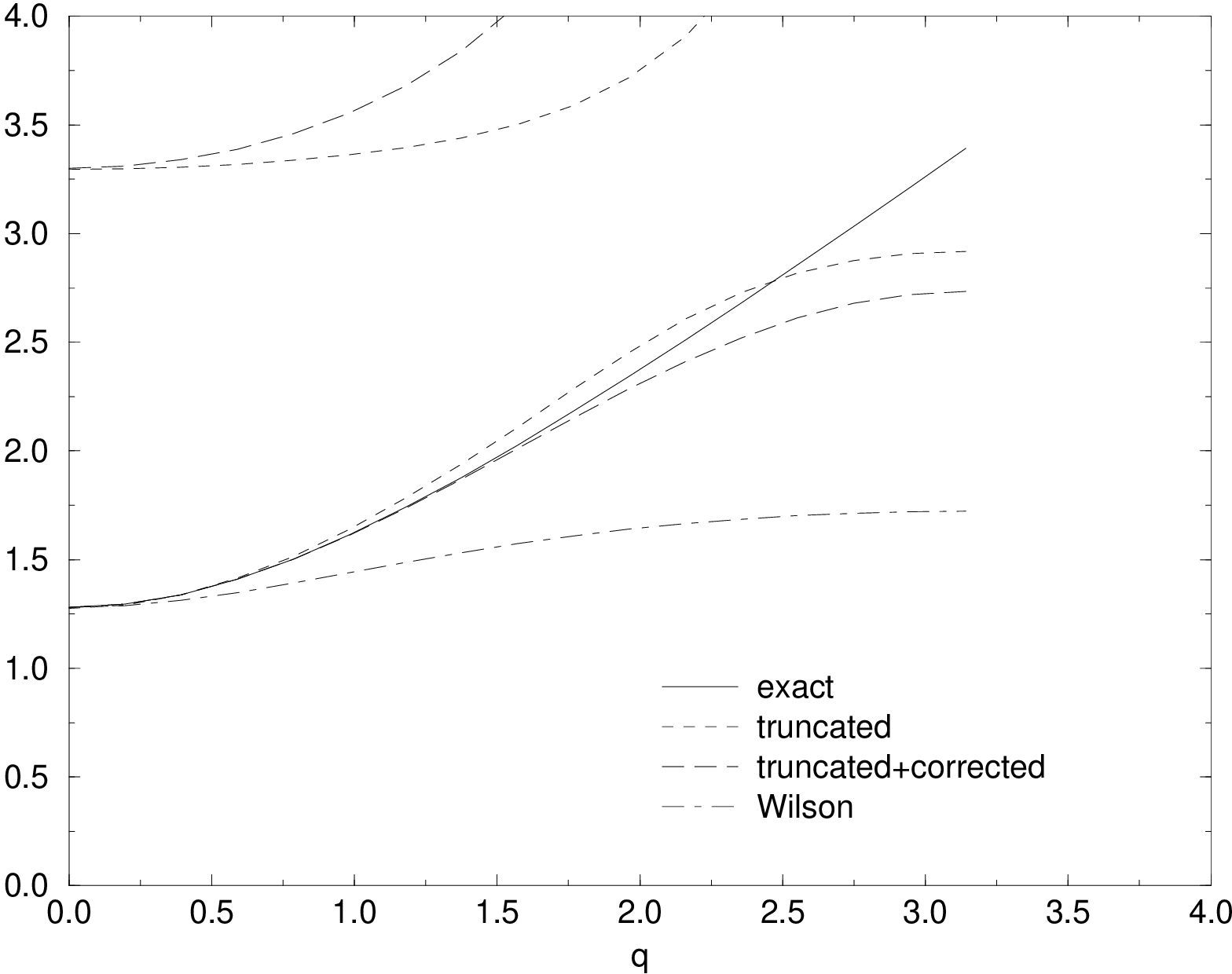}
\vspace{-10pt}
\caption{Fermion spectrum at $ma=1.28$.}
\label{fig:sp1280}
\end{figure}

Fig.~\ref{fig:sp0005} shows the spectrum $E(q,0,0)$ (along the axis)
at $ma=0.005$ (for the fermionic action with this small mass
and corresponding to the same RGT as above),
as a typical example. The solid line is the exact spectrum, 
the dashed is for the action truncated to the hypercube, 
the long--dashed is the truncated one with tiny corrections
to satisfy the Symanzik condition, the dot--dashed is for
the Wilson  fermions.
Fig.~\ref{fig:sp1280} shows the same quantities for $m=1.28$.
(Note that the Fermilab action \cite{Alford} on the 
{\it isotropic} lattice gives a far worse dispersion relation: 
for $m=0$ it is complex and for $ma>0$ the kinetic mass is too small,
see a figure in Ref.~\cite{Wiese}. 
Remember, however, that it is assumed to be used on anisotropic 
lattices.)

The spectrum of the full FP action is, of course, exact for any
value of the fermion mass (or rather $ma$). 
The coefficients $\rho_{\mu}$ and $\lambda$ in eq.~(\ref{fermfp})
depend on the mass, and they decay exponentially in the distance 
$r$ as  $\propto \exp(-\gamma r)$.
A truncation at some small distance should necessarily induce 
`ghosts' in the spectrum (higher branches, which are artifacts) 
with an energy $\sim \gamma$. A truncation is sensible therefore
only if $ma < \gamma$. It is not yet clear what is the largest
mass this condition allows ($\gamma$ will also depend
on $ma$), but $ma\approx 2$ is still definitely allowed. 
To study higher lattice masses, one might need to consider
anisotropic lattices, anyhow.

\begin{figure}[htb]
\vspace{9pt}
\epsfxsize=62mm
\hspace{6mm}
\epsfbox{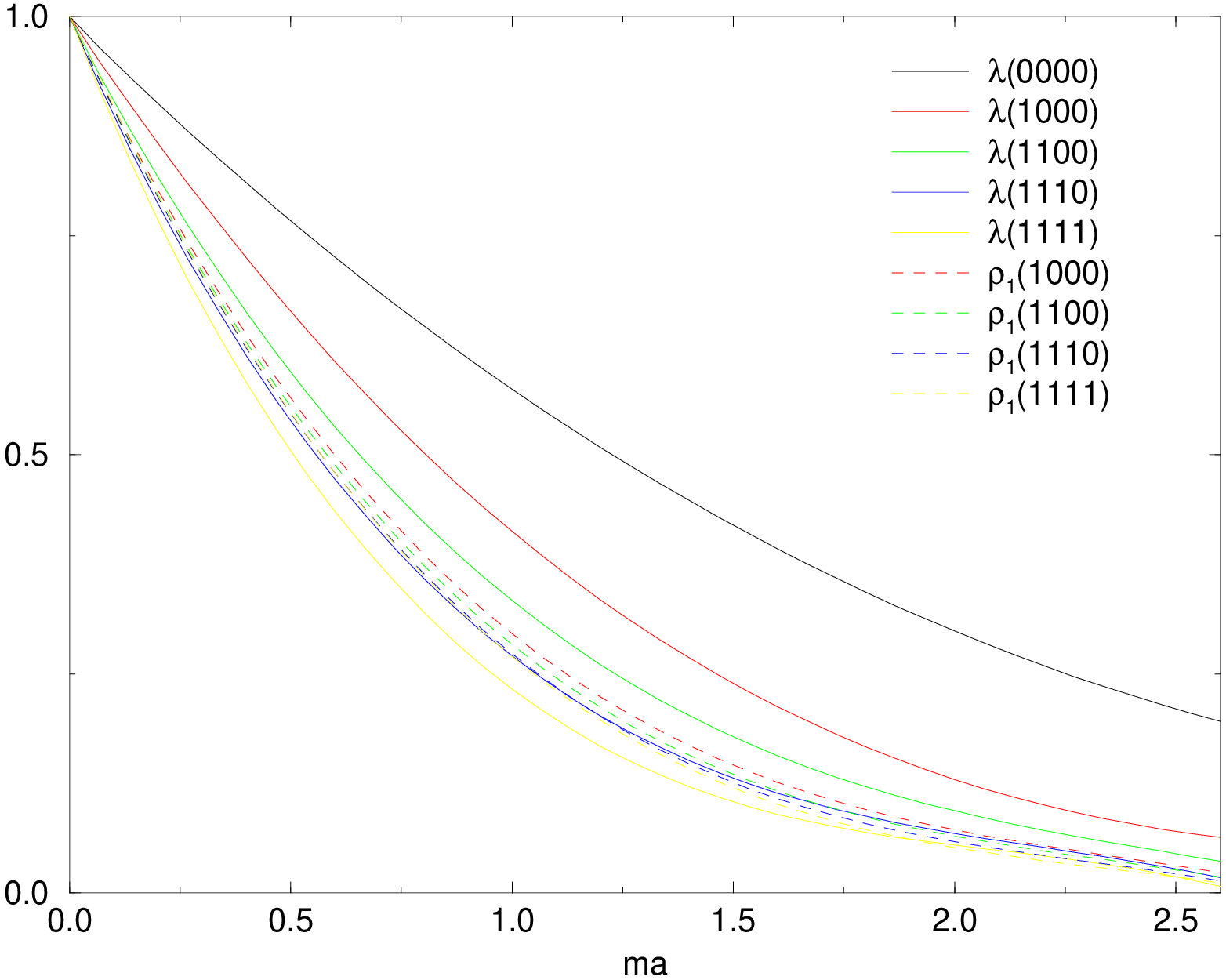}
\vspace{-10pt}
\caption{Mass dependence of the ratios $\lambda(r,ma)/\lambda(r,0)$
and $\rho_1(r,ma)/\rho_1(r,0)$.}
\label{fig:ratios}
\end{figure}

The mass dependence of the coefficients in eq.~(\ref{fermfp})
follows a smooth curve which could be parametrized easily.
In Fig.~(\ref{fig:ratios}) the ratios $\lambda(r,ma)/\lambda(r,0)$ 
and $\rho_1(r,ma)/\rho_1(r,0)$ are shown for different $r$ values 
on the hypercube, only as an illustration to this statement.
(Note that this dependence is obtained by an optimized 
`running RGT'.)

\begin{figure}[htb]
\vspace{9pt}
\epsfxsize=62mm
\hspace{6mm}
\epsfbox{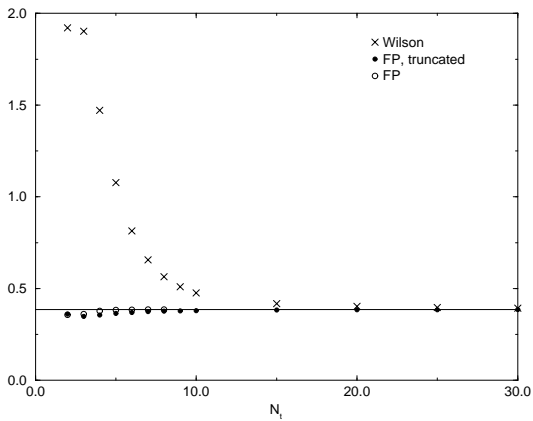}
\vspace{-10pt}
\caption{$p/T^4$ vs. $N_t$ for massless fermions.}
\label{fig:fthermo}
\end{figure}

As we saw, thermodynamic quantities are very sensitive to the cut--off
effects. In Fig.~\ref{fig:fthermo} for the case of massless fermions
the ratio $p/T^4$ is plotted vs. $N_t$, for the Wilson action,
the full FP action and the FP action truncated to the unit hypercube
(but still uncorrected for the tiny violation of the Symanzik
condition). Note that the full FP action has a correction vanishing
exponentially with increasing $N_t$.

\subsection{Fermions and gauge fields}

The final goal of the improvement approach is to produce a good
action of the full QCD. This goal is not achieved yet, but some 
important steps are already made.
I shall review here the situation within the RG approach.
To derive the FP action for the case of fermions~+~gluons
the following steps have to be performed:
\begin{enumerate}
\item
Find the FP action for the pure gauge fields and free fermions,
eqs.~(\ref{fpeq},\ref{fermfp}),
and obtain the corresponding minimized fine fields from the coarse
ones, $U=\overline{U}(V)$, $\psi=\psi(\Psi)$, etc.

\item
Substitute this into the interaction part in the saddle point
equation:
\end{enumerate}
\beea
& & S'_{\rm F}(\bar{\Psi},\Psi,V) =
S_{\rm F}(\bar{\psi},\psi,U) +  \\
& & ~
\left. T_{\rm F}(\bar{\Psi},\Psi,V,\bar{\psi},\psi,U)
\right|_{U=\overline{U}(V),\psi=\psi(\Psi),\ldots} \nonumber
\eea

This equation is already satisfied for $V\equiv 1$
since it is the FP equation for the free fermionic action.
For smooth fields $V$ one can solve it by iteration,
obtaining the linear (in $A_{\mu}(x)$) approximation to the 
interaction vertex. 
This can be expressed in the gauge invariant form
\bee
\bar{\psi}(n) \Gamma_A R_A(n,n',U) \psi(n')
\label{linappr}
\ee
where $\Gamma_A$ is a Dirac matrix and $R_A(n,n',U)$ is a sum 
of weighted paths connecting the sites $n$ and $n'$.
For the MIT blocking this approximation has been done
in Ref.~\cite{BW}. 
For the Bern/Boulder blocking(s) the details have not been 
published yet, but the main points are presented at this
conference \cite{DG_talk}.
The interaction vertex $R_A(n,n',U)$ in the linear approximation
looks roughly as follows. For $\Gamma=1$ or $\gamma_{\mu}$
the coefficients are obtained by gauging the free fermion FP
action: the nearest neighbours are connected by the direct link and
the 6 staples with a weight of 50--50\%, the others on the unit
hypercube predominantly by the average of the corresponding 
shortest paths.
The $\Gamma=\sigma_{\mu\nu}$ term (not present in the
free fermion action) is rather small, the $\gamma_5$ and
$\gamma_{\mu}\gamma_5$ terms are even smaller.

The MIT group made a preliminary study \cite{Wiese} in the quenched 
approximation --- using a naive gauging of the free fermionic
FP action, they measured the pion dispersion relation with 
zero bare quark mass, at $\beta=5$. Instead of a light pion
they obtained a heavy one ($m_{\pi}a\approx 3$), but with a 
surprizingly good dispersion relation.

Note that based on the 1--loop perfectness, for the full QCD 
FP action we do not expect a large perturbative renormalization.
This should be, however, the consequence of terms non--linear
in the link variables $U$ in the interaction vertex
$R_A(n,n',U)$ (cf. eq.~(\ref{Sparam})). 
Only after including these non--linear terms should one expect
a small bare mass renormalization.

\section{CONCLUSION}

The size of the lattice artifacts depends on the model investigated,
on the particular form of the action used, on the quantity one
measures, and, of course, on the actual lattice spacing $a$.
The answer to the question that which improved action should be 
preferred depends also on many factors --- in particular the 
overhead in the MC simulations, the expected improvement, etc.
Since the computing costs grow like a (sometimes large) power of the
inverse lattice spacing, even a large overhead can be tolerated if
the corresponding artifacts are sufficiently suppressed.

At  lattice spacings used in present numerical simulations
the tree--level or one--loop Symanzik improved actions do not
seem to improve sufficiently. One solution is to modify
them in a non--perturbative way. Both the tadpole improvement
and the method of non--perturbative determination of the improvement
coefficients go in this way. The actions used are relatively simple
modifications of the standard action, hence the overhead is 
relatively low.

The more ambitious RG approach has a longer incubation period ---
it is not easy to derive and optimize the FP actions. 
Since their structure is more complicated the overhead is larger.
The experience collected so far indicates, however, that the
resulting artifacts are very much suppressed, and it might
compensate the extra complications of this approach.
As mentioned, the truncation of the FP action can be made to 
preserve the Symanzik conditions, so it suggests a natural choice
in the Symanzik improvement program. In measuring the topological
susceptibility this approach has the advantage to represent
properly the classical instantons, avoiding the main problems
of the standard approach. One can speculate that in cases
where topological effects have strong influence on the behaviour
of the system, the use of the FP actions is preferable
in spite of the extra overhead.

For spin systems, pure gauge theories and free fermions there
exist good choices of improved actions, which have passed
various tests.  For full QCD at present there are only promising 
candidates --- in the coming year(s) the situation will be,
hopefully, further improved.

I would like to thank Anna Hasenfratz, Peter Hasenfratz and 
Peter Weisz for useful comments.

This work was supported by the Swiss National Science Foundation.

\end{document}